\documentclass[preprint,twocolumn,3p,lefttitle,sort&compress]{elsarticle}
\usepackage{epsfig}
\usepackage{amsmath,amssymb,amsfonts}
\usepackage{mathrsfs}
\usepackage{bbm}
\usepackage{slashed}
\usepackage{graphicx}
\usepackage{verbatim}
\usepackage[T1]{fontenc}
\usepackage[utf8]{inputenc}
\usepackage[colorlinks]{hyperref}
\usepackage{ifthen}
\usepackage{forloop}
\usepackage[english]{babel}
\usepackage{mathbbol}
\usepackage[usenames]{color}

\newcommand{\FRG}{{\small FRG}}
\newcommand{\RG}{{\small RG}}

\newcommand{\CDT}{{\small CDT}}

\newcommand{\SM}{{\small SM}}
\newcommand{\HL}{{\small HL}}
\newcommand{\eg}{{\textit{e.g.}}}
\newcommand{\ie}{{\textit{i.e.}}}

\begin{document}

\title{Lorentz symmetry is relevant}
\author{Benjamin Knorr}
\ead{b.knorr@science.ru.nl}
\address{Institute for Mathematics, Astrophysics and Particle Physics (IMAPP), Radboud University Nijmegen, Heyendaalseweg 135, 6525 AJ Nijmegen, The Netherlands}

\begin{abstract}
We set up a covariant renormalisation group equation on a foliated spacetime which preserves background diffeomorphism symmetry.
As a first application of the new formalism, we study the effect of quantum fluctuations in Lorentz symmetry breaking theories of quantum gravity.
It is found that once a small breaking is introduced e.g. at the Planck scale, quantum fluctuations enhance this breaking at low energies.
A numerical analysis shows that the magnification is of order unity for trajectories compatible with a small cosmological constant.
The immediate consequence is that the stringent observational constraints on Lorentz symmetry breaking are essentially scale-independent and must be met even at the Planck scale.
\end{abstract}
%

%
\maketitle
\section{Introduction}

The beginning of the 20th century has been a very successful time for theoretical physics. On the one hand, the foundations of quantum mechanics were laid out, leading ultimately to the formulation of the Standard Model (\SM{}) which describes the electromagnetic, weak and strong interactions. On the other hand, Einstein put forward his theory of General Relativity, showing that gravity could be seen as the effect of the curvature of spacetime. Despite tremendous effort, no consistent theory combining quantum mechanics with gravity is available at present. Different contesters include Loop Quantum Gravity \cite{Thiemann:2002nj, Rovelli:2004tv}, String Theory \cite{Becker:2007zj, Schomerus:2017lqg}, Asymptotic Safety, both in continuum \cite{Niedermaier:2006wt, Reuter:2012id, Percacci:2017fkn, Eichhorn:2017egq} and discrete \cite{Ambjorn:2012jv} formulations, Causal Sets \cite{Sorkin:2003bx, Surya:2011yh} and many more. Nevertheless, none of the approaches can claim full success in the combination of gravity and the \SM{}.

A key guiding principle in the construction of the \SM{} is Lorentz invariance. Experimentally this is well justified: we do not have any reason to believe that at the presently available energies this symmetry is broken \cite{Mattingly:2005re, Liberati:2013xla}. However, it is well-known that coupling constants in quantum field theories as the \SM{} depend on the energy scale of the process under consideration. It is thus conceivable that coupling constants related to Lorentz-breaking terms are important at trans-Planckian energy scales, whereas they become negligible at scales currently accessible by experiment. This is the general idea of Ho\v{r}ava-Lifshitz (\HL{}) quantum gravity \cite{Horava:2008ih, Horava:2009uw}. In this approach one tries to circumvent the problem of the perturbative non-renormalisability of quantum gravity by introducing an anisotropic scaling in the form of higher order spatial derivatives of the metric. These might cure the ultraviolet divergences and thus make a perturbative quantisation of gravity similar to that of the \SM{} possible \cite{Barvinsky:2015kil, Barvinsky:2017kob}. For reviews of \HL{} gravity, see \cite{Mukohyama:2010xz, Sotiriou:2010wn, Wang:2017brl}, and for the experimental status see \cite{Mattingly:2005re, Kostelecky:2008ts, Blas:2012vn, Liberati:2013xla, Audren:2014hza, Yagi:2013qpa, Yagi:2013ava, Blas:2014aca, Hees:2016lyw}.

An open issue in \HL{} gravity is whether Lorentz invariance can be restored, at least to a high precision, at large enough length scales. Depending on whether quantum fluctuations enhance or diminish the breaking, \HL{} is viable as a theory of quantum gravity. In this letter, we will analyse this question by setting up a renormalisation group (\RG{}) equation for foliated spacetimes which reduces to the equation derived in a covariant setting if no breaking terms are present.

\section{Functional renormalisation group}

The functional renormalisation group (\FRG{}) is a versatile non-perturbative tool to investigate quantum fluctuations. We will use the formulation introduced in \cite{Wetterich:1992yh, Morris:1993qb, Reuter:1996cp},
\begin{equation}\label{eq:flow}
 k \partial_k \Gamma_k = \frac{1}{2} \text{STr} \left[ \left( \Gamma^{(2)}_k + \mathfrak R_k \right)^{-1} k \partial_k \mathfrak R_k \right] \, .
\end{equation}
In this equation, $\Gamma_k$ is the effective average action which describes processes at some scale $k$, $\Gamma^{(2)}_k$ its second functional derivative with respect to the dynamical fields, and $\mathfrak R_k$ is a regulator which acts as a momentum-dependent mass term. Finally, the supertrace STr sums over discrete and integrates over continuous indices. Due to the regularisation, the equation is well-defined both in the ultraviolet and the infrared.

This equation has seen successful exploitation in a variety of contexts in quantum field theory and condensed matter, but in particular also in quantum gravity \cite{Christiansen:2014raa, Gies:2015tca, Demmel:2015oqa, Ohta:2015efa, Ohta:2015fcu, Christiansen:2015rva, Meibohm:2015twa, Henz:2016aoh, Gies:2016con, Morris:2016spn, Denz:2016qks, Falls:2016msz, Meibohm:2016mkp, Eichhorn:2016vvy, Gonzalez-Martin:2017gza, Christiansen:2017bsy, Christiansen:2017gtg, Becker:2017tcx, Christiansen:2017cxa, Eichhorn:2017ylw, Eichhorn:2017lry, Falls:2017lst, Eichhorn:2017sok, Platania:2017djo, Eichhorn:2017muy, Knorr:2017fus, Knorr:2017mhu, Eichhorn:2018akn, Alkofer:2018fxj, Gies:2018jnv, Gubitosi:2018gsl, Alkofer:2018baq, Eichhorn:2018ydy}. For reviews of the \FRG{} in quantum gravity see \cite{Reuter:1996ub,Pawlowski:2005xe,Niedermaier:2006wt,Percacci:2007sz,Reuter:2012id,Nagy:2012ef,EichhornReview18}, and for a recent reformulation taking appropriate care of the normalisation of the partition function see \cite{Lippoldt:2018wvi}.

The technical challenge of the present work is to implement this flow equation for quantum gravity in a foliated setup. Earlier approaches \cite{Manrique:2011jc, Rechenberger:2012dt, Contillo:2013fua, DOdorico:2014tyh, Biemans:2016rvp, Biemans:2017zca, Houthoff:2017oam, Platania:2017djo} have not been able to retain the full background diffeomorphism invariance for invariant theories, and thus in these approaches it is hard to decide whether contributions to the energy dependence of symmetry-breaking couplings stem from these breaking terms or from genuine physical effects. In the following, we will advocate a formulation without this deficit.

\section{Foliation setup}

Renormalisation group calculations in quantum gravity heavily rely on the use of the background method. In this, the dynamical, $d$-dimensional (Lorentzian) metric $g_{\mu\nu}$ is split into a fixed but arbitrary background metric $\bar g_{\mu\nu}$, and (not necessarily small) perturbations $h_{\mu\nu}$ around that. This entails, for a linear parameterisation of the perturbation,
\begin{equation}
 g_{\mu\nu} = \bar g_{\mu\nu} + h_{\mu\nu} \, .
\end{equation}
Nonlinear parameterisations have received a lot of interest recently \cite{Nink:2014yya, Demmel:2015zfa, Percacci:2015wwa, Gies:2015tca, Labus:2015ska, Ohta:2015zwa, Ohta:2015efa, Falls:2015qga, Ohta:2015fcu, Dona:2015tnf, Ohta:2016npm, Falls:2016msz, Percacci:2016arh}.
Let us now establish the foliation setup. For the given Lorentzian metric $g_{\mu\nu}$, we introduce a timelike, normalised vector $n$ and a spatial metric $\sigma$ orthogonal to $n$, such that
\begin{equation}\label{eq:foliation}
\begin{aligned}
 g_{\mu\nu} &= \sigma_{\mu\nu} + n_\mu n_\nu \, , \\
 g^{\alpha\beta}n_\alpha \sigma_{\beta\gamma} &= 0 \, , \\
 g^{\mu\nu}n_\mu n_\nu &= 1 \, .
\end{aligned}
\end{equation}
For the study of the renormalisation of local actions, it is presumably enough to assume local existence of the vector field $n_\mu$, and in the following we will restrict ourselves to this case.

The central idea of this work is to start in a background diffeomorphism invariant setup in the metric language, and then to replace the metric perturbation $h$ by the corresponding foliation perturbations, restricting the path integral to foliated spacetimes only. This is most easily done if this map is linear, as then the one-loop structure of the flow equation \eqref{eq:flow} is preserved. For this reason, we choose a linear parameterisation for the $n$-fluctuations, but a particular quadratic parameterisation for the $\sigma$-fluctuations:
\begin{equation}
\begin{aligned}
 n_\mu &= \bar n_\mu + \hat n_\mu \, , \\
 \sigma_{\mu\nu} &= \bar \sigma_{\mu\nu} + \hat\sigma_{\mu\nu} - \hat n_\mu \hat n_\nu \, .
\end{aligned}
\end{equation}
Hatted quantities refer to the fluctuating fields. We also assume that the background quantities fulfil the corresponding relations of normalisation and orthogonality \eqref{eq:foliation}. With this parameterisation, the map between $h$ and the foliation fluctuations is indeed linear as the quadratic piece cancels,
\begin{equation}\label{eq:h-fol-map}
 h_{\mu\nu} = \hat \sigma_{\mu\nu} + \bar n_\mu \hat n_\nu + \hat n_\mu \bar n_\nu \, .
\end{equation}
Clearly, in the process of going from $h$ to $\hat \sigma$ and $\hat n$, we increased the number of degrees of freedom from 10 (symmetric matrix in 4d) to 14 (symmetric matrix plus vector). On the other hand, $\hat\sigma$ and $\hat n$ are not completely independent, as the full spatial metric $\sigma$ and timelike vector $n$ have to fulfil their respective constraints \eqref{eq:foliation}. A short calculation shows that the simplest solution to both constraints is\footnote{More complicated solutions to the constraints exist which differ by terms at least quadratic in the fluctuations. These do not contribute in the subsequent approximations, where only the linear part is important; other choices of solutions will be considered elsewhere.}
\begin{equation}
 \mathfrak F_\nu := \bar n^\mu \hat \sigma_{\mu\nu} - \bar n^\mu \hat n_\mu \hat n_\nu = 0 \, .
\end{equation}
This constraint is implemented via a Lagrange multiplier, similar to a gauge fixing, and we will call this procedure suggestively foliation gauge fixing. No (dynamical) ghosts arise from this procedure as the functional $\mathfrak F_\nu$ is ultralocal.

With these two ingredients, we already have the dictionary between the metric and the foliation language. On a path integral level,
\begin{equation}\label{eq:pi}
\begin{aligned}
 Z &\sim \! \! \int \! \! \mathcal D h \, e^{\mathbf{i} (S[\bar g,h] + S_\text{gf}[\bar g,h])} \\
 &\sim \! \! \int \! \! \mathcal D \hat \sigma \mathcal D \hat n \, e^{\mathbf{i} (S[\bar g,\hat \sigma, \hat n] + S_\text{gf}[\bar g, \hat \sigma, \hat n] + S_\text{f}[\bar g, \hat \sigma, \hat n])} \, ,
\end{aligned}
\end{equation}
where $S$ is some gravitational action, \eg{} the Einstein-Hilbert action or an $f(R)$ action.
Let us stress at this point that in the path integral we only include foliatable spacetimes.
Moreover,
\begin{equation}\label{eq:fgf}
 S_\text{f} = \frac{1}{32\pi G_N \alpha_\text{fol}} \int \sqrt{|\bar g|} \bar g^{\mu\nu} \mathfrak F_\mu \mathfrak F_\nu
\end{equation}
is the foliation gauge fixing action with foliation gauge parameter $\alpha_\text{fol}$, $S_{gf}$ is the gauge fixing action, and we suppressed the integral over the Faddeev-Popov ghosts and their corresponding action. For simplicity, we will choose a harmonic gauge fixing,
\begin{equation}\label{eq:gf}
\begin{aligned}
 S_\text{gf} &= \frac{1}{32\pi G_N} \int \sqrt{|\bar g|} \bar g^{\mu\nu} \mathcal F_\mu \mathcal F_\nu \, , \\
 \mathcal F_\mu &= \left( \delta^\alpha_\mu \bar D^\beta - \frac{1}{2} \bar g^{\alpha\beta} \bar D_\mu \right) h_{\alpha\beta} \, .
\end{aligned}
\end{equation}
For a recent analysis of the gauge dependence of the renormalisation group behaviour of quantum gravity, we refer the reader to \cite{Gies:2015tca, Knorr:2017fus}.

\subsection{Approximations}

Having specified this setup, we can now apply the standard machinery of the \FRG{} to obtain the renormalisation group running of couplings appearing in a given action. Since we have access to a foliation structure, our action can include terms which break the full diffeomorphism symmetry but are invariant under foliation preserving diffeomorphisms. In the following, we will combine the Einstein-Hilbert action,
\begin{equation}\label{eq:EH}
 S_\text{EH} = \frac{1}{16\pi G_N} \int \sqrt{|g|} \left( -R + 2 \Lambda \right) \, ,
\end{equation}
with Newton's constant $G_N$ and the cosmological constant $\Lambda$ as the two coupling constants, and all breaking terms with up to two derivatives,
\begin{equation}\label{eq:Sbreak}
 \tilde S \! = \! \frac{1}{16\pi G_N} \! \! \int \! \! \sqrt{|g|} \left( k_2 K^{\mu\nu} K_{\mu\nu} \! + \! k_0 K^2 \! + \! a_1 \mathcal A^\mu \mathcal A_\mu \right) \, ,
\end{equation}
with breaking coupling constants $k_0, k_2$ and $a_1$.
In this,
\begin{equation}
 K_{\mu\nu} = \frac{1}{2} \left( n^\alpha D_\alpha \sigma_{\mu\nu} + D_\mu n_\nu + D_\nu n_\mu \right)
\end{equation}
is the extrinsic curvature of spatial slices, and is orthogonal to the normal vector,
\begin{equation}
 n^\mu K_{\mu\nu} = 0 \, .
\end{equation}
$K$ is the trace of the extrinsic curvature,
\begin{equation}
 \mathcal A_\mu = n^\alpha D_\alpha n_\mu
\end{equation}
is the so-called acceleration vector. A term proportional to the intrinsic (3-dimensional) Ricci scalar can be reabsorbed in the terms already present by a Gauss-Codazzi equation, up to a total derivative which we neglect. Our ansatz for the action appearing in the path integral is thus
\begin{equation}\label{eq:ansatz}
 S = S_\text{EH} + \tilde S \, .
\end{equation}

Before we go on and derive the \RG{} running of the 5 couplings $(G_N,\Lambda,k_0,k_2,a_1)$, let us emphasise that the given construction links the fully diffeomorphism invariant metric language on foliatable spacetimes to a still invariant language which has explicit access to the foliation structure. This construction is intimately related to the question of a well-defined Wick rotation. A closely related proposal for a Wick rotation in curved spacetimes can be found in \cite{PhysRevD.15.1494}, which should carry over to the present setup. Thus this flow provides a link for previous Euclidean \RG{} studies of quantum gravity to flows on foliatable Lorentzian spacetimes. The \RG{} flow of the non-breaking couplings in the foliation language \emph{is the same} as in the metric language. The new implementation is in spirit very close to causal dynamical triangulations (\CDT{}), and the present \RG{} equation allows a more direct connection of the flow in the discrete and the continuum.

\section{Foliated renormalisation group equation}

We are now in the situation to set up the \RG{} flow for foliated spacetimes. In the following, we will use the background field approximation: once the hessian $\Gamma^{(2)}_k$ is calculated, the fluctuations are set to zero. This is for technical simplicity, for more elaborate approximations retaining parts of the fluctuation dependence in pure gravity see \cite{Christiansen:2012rx, Codello:2013fpa, Christiansen:2014raa, Christiansen:2015rva, Dona:2015tnf, Meibohm:2015twa, Denz:2016qks, Meibohm:2016mkp, Henz:2016aoh, Christiansen:2016sjn, Knorr:2017fus, Christiansen:2017cxa, Christiansen:2017bsy, Knorr:2017mhu, Eichhorn:2018akn}. To our ansatz for the effective action \eqref{eq:ansatz}, where all couplings are replaced by $k$-dependent counterparts, we add the standard and foliation gauge fixings, \eqref{eq:gf} and \eqref{eq:fgf}. Finally, we have to specify the regulator. For this, we take
\begin{equation}
 \Delta S_k = \frac{1}{2} \frac{1}{16\pi G_N} \int \sqrt{|\bar g|} \, h \left[  \mathbbm 1 - 2 \Pi_\text{tr} \right] \mathfrak R_k(\bar\Delta) \, h \, ,
\end{equation}
where $\bar\Delta = -\bar D^2$ is the background covariant Laplacian, $\mathbbm 1$ is the unit operator for symmetric tensors, $\Pi_\text{tr}$ is the projector onto the trace, $\mathfrak R_k(\bar\Delta)$ is the regulator function and $h$ is understood to be replaced by the foliation fluctuations according to \eqref{eq:h-fol-map}. Clearly, this regulator preserves background diffeomorphism invariance by construction. On the other hand, in foliation language, nontrivial curvature terms are included in exactly the right way to preserve the symmetry. These additional terms would be very hard to guess if a regulator would be constructed directly for the foliation fluctuations. In the ghost sector, we use a similar standard regularisation. In the background field approximation, the ghost contribution towards the flow is anyway the same as in the non-foliated setup. Since it does not involve graviton fluctuations, it can be directly copied from the literature \cite{Reuter:2001ag}.

The actual calculation of the \RG{} flow is implemented by the Mathematica package \textit{xAct} \cite{xActwebpage, Brizuela:2008ra, 2008CoPhC.179..597M, 2007CoPhC.177..640M, 2008CoPhC.179..586M, 2014CoPhC.185.1719N}. In the evaluation of the trace, one additional advantage of the present covariant approach manifests itself: we can employ standard heat kernel techniques, see \eg{} \cite{Vassilevich:2003xt, Groh:2011dw}, and do not have to resort to the much more complicated heat kernel for anisotropic operators \cite{Nesterov:2010yi, DOdorico:2015pil, Barvinsky:2017mal}. Nevertheless, due to the additional background foliation structure, the technical complexity is significantly larger than for unfoliated calculations. For that reason, we will restrict ourselves to the most interesting part of the flow: we only consider the diffeomorphism invariant part plus terms linear in the breaking couplings. With this, we can already evaluate whether Lorentz symmetry-breaking theories of quantum gravity stand a chance in restoring the symmetry at low energies. Let us finally note that to calculate the heat kernel trace, we assume that we can rotate to Euclidean signature.

\section{Results}

For convenience let us first introduce the threshold integrals
\begin{equation}
\begin{aligned} 
 Q_{n,m}^\alpha(\mu) &:= \!\! \int_0^\infty \!\!\!\! \text{d}z \tfrac{z^n \left((2 - \alpha) \mathfrak R_k(z) - 2 z \mathfrak R_k'(z)\right)}{\left(z+\mathfrak R_k(z)+\mu\right)^m} \, , \\
 \tilde Q_{n,m}^\alpha(\mu) &:= \!\! \int_0^\infty \!\!\!\! \text{d}z \tfrac{z^n \left((2 - \alpha) \mathfrak R_k(z) - 2 z \mathfrak R_k'(z)\right)\left(1+\mathfrak R_k'(z)\right)^2}{\left(z+\mathfrak R_k(z)+\mu\right)^m} .
\end{aligned}
\end{equation}
We furthermore introduce the dimensionless couplings
\begin{equation}
 g = G_N k^2 \, , \, \lambda = \Lambda/k^2 \, ,
\end{equation}
and an overdot shall indicate a $(k\partial_k)$-derivative.
The graviton anomalous dimension is then given by
\begin{equation}
 \eta = \frac{\dot g - 2 g}{g} \, .
\end{equation}
With this, we can write down the flow equations for our system:
\begin{equation}\label{eq:gdot}
\begin{aligned} 
 -\frac{\eta}{16\pi g} &= -\frac{5}{96\pi^2} Q_{0,1}^\eta(-2\lambda) + \frac{1}{16\pi^2} Q_{1,2}^0(0) \\
 &\quad + \frac{1}{24\pi^2} Q_{0,1}^0(0) + \frac{3}{16\pi^2} Q_{1,2}^\eta(-2\lambda) \\
 &\quad + \frac{13a_1-3k_0+9k_2}{768\pi^2} Q_{1,2}^\eta(-2\lambda) \\
 &\quad - \frac{11a_1+k_0+7k_2}{128\pi^2} Q_{2,3}^\eta(-2\lambda)  \, ,
\end{aligned}
\end{equation}
\begin{equation}\label{eq:lambdadot}
\begin{aligned} 
 \frac{\dot\lambda+(2-\eta)\lambda}{8\pi g} &= \frac{5}{16\pi^2} Q_{1,1}^\eta(-2\lambda) - \frac{1}{4\pi^2}Q_{1,1}^0(0) \\
 &\quad - \frac{3(5a_1-k_0+3k_2)}{256\pi^2} Q_{2,2}^\eta(-2\lambda) \, ,
\end{aligned}
\end{equation}
\begin{equation}\label{eq:k0dot}
\begin{aligned} 
 \frac{\dot k_0-\eta k_0}{16\pi g} &= -\frac{22a_1-69k_0-3k_2}{384\pi^2} Q_{1,2}^\eta(-2\lambda) \, ,
\end{aligned}
\end{equation}
\begin{equation}\label{eq:k2dot}
\begin{aligned} 
 \frac{\dot k_2-\eta k_2}{16\pi g} &= \frac{22a_1-3k_0+39k_2}{384\pi^2} Q_{1,2}^\eta(-2\lambda) \\
 &\quad + \frac{16a_1+3k_0+k_2}{384\pi^2} \tilde Q_{3,4}^\eta(-2\lambda) \, ,
\end{aligned}
\end{equation}
\begin{equation}\label{eq:a1dot}
\begin{aligned} 
 \frac{\dot a_1-\eta a_1}{16\pi g} &= -\frac{a_1+2k_0+4k_2}{32\pi^2} Q_{1,2}^\eta(-2\lambda) \\
 &\quad + \frac{16a_1+3k_0+k_2}{128\pi^2} \tilde Q_{3,4}^\eta(-2\lambda) \, .
\end{aligned}
\end{equation}
By construction, the flow equations of the breaking couplings vanish when the breaking couplings themselves vanish, and the equations for $\dot g$ and $\dot \lambda$ reduce to the flow equations of the covariant setting \cite{Reuter:2001ag}. Notice that the equations are independent of $\alpha_\text{fol}$, which is due to the linearisation in the breaking couplings. It comes about by the fact that the foliation gauge fixing operator only appears in the $\hat\sigma\hat\sigma$-part of the hessian, is proportional to $\bar n$, and all propagators (which are expanded in the breaking couplings) are, to linear order in the breaking, contracted with the regulator, which is proportional to $\bar\sigma$ in that sector. Terms of higher order in the breaking couplings are expected to be foliation gauge dependent. The set of equations \eqref{eq:gdot}-\eqref{eq:a1dot} constitutes one of the main results of this work.

To further analyse the flow equations, we will use the Litim regulator \cite{Litim:2001up, Litim:2002cf},
\begin{equation}
 \mathfrak R_k(z) = (1-z)\theta(1-z) \, ,
\end{equation}
where $\theta$ is the Heaviside theta function. We further linearise the flow of the breaking couplings in all couplings (not counting the overall prefactor of $g$), which simplifies the subsequent discussion and gives the leading order behaviour near the diffeomorphism symmetric hypersurface in coupling space. This reduces the equations to
\begin{equation}\label{eq:lineqs}
\begin{aligned}
 \dot k_0 &= -\frac{g}{24\pi} (22a_1 + 19k_0 - 3k_2) \, , \\
 \dot k_2 &= \frac{g}{24\pi} (22a_1 - 3k_0 - 49k_2) \, , \\
 \dot a_1 &= -\frac{g}{6\pi} (25a_1+6k_0+12k_2) \, .
\end{aligned}
\end{equation}
Notice that once any of the couplings is present, it immediately generates the other couplings. This implies that reductions to subsets, \eg{} the so-called $\lambda-R$ model \cite{Horava:2009uw, Bellorin:2010je, Loll:2014xja, Loll:2017utf, Pires:2018srq} which only retains $k_0$, are in general not stable under renormalisation.

Before analysing the equations, let us point out their range of viability. Clearly, for large breaking couplings, the linear approximation is not applicable, we are thus confined to the situation of small breaking. We also expanded in $g$ and $\lambda$, thus they should be small, which is the case in the semi-classical regime\footnote{Note that in the deep infrared, the dimensionless cosmological constant goes to $\infty$ owing to a finite dimensionful cosmological constant. Since $\lambda$ appears in denominators only, this further suppresses the flow, which is already suppressed by a dimensionless Newton's constant which goes to zero. The approximation thus should do fine even in the deep infrared as long as we do not take into account positive powers of $\lambda$.}. Here we assume that the flow is close to the Gaussian fixed point, in agreement with observations which find a very small cosmological constant \cite{Reuter:2004nx}. From previous studies in the field, one sees the generic feature that at energies just below the Planck scale, the couplings already run classically, see \eg{} \cite{Reuter:2001ag, Christiansen:2014raa, Denz:2016qks, Gubitosi:2018gsl}. Thus we expect the equations to be valid from slightly below the Planck scale to the regime where the dimensional running of the cosmological constant sets in. This should include most of the phenomenologically interesting scales.

We can now make a statement about dynamical symmetry restoration. Assume that Lorentz symmetry is broken by some small amount, parameterised by the couplings $(k_0,k_2,a_1)$, at some high energy scale $\Lambda_\text{UV}$. Using \eqref{eq:lineqs}, one sees that for a fixed value of $g>0$, the flow points towards the origin. To see this, we take the scalar product of the radial vector $(k_0,k_2,a_1)$ with the vector field $(\dot k_0, \dot k_2,\dot a_1)$,
\begin{equation}
\begin{aligned}
 &(k_0,k_2,a_1) \cdot (\dot k_0, \dot k_2,\dot a_1) \\
 &\quad= -\frac{g}{24\pi} \bigg[ \frac{23}{2}(2a_1+k_0)^2 + \frac{13}{2}(2a_1+k_2)^2 \\
 &\quad\quad\quad\quad\quad\quad+ 28 a_1^2 + \frac{15}{2} k_0^2 + \frac{85}{2} k_2^2 \bigg] \leq 0 \, ,
\end{aligned}
\end{equation}
which is non-positive, \ie{} it points inwards. This means that if we \textit{decrease} the energy scale (\textit{increase} the length scale), the breaking couplings \textit{grow} generically. The consequence of this is that generically, quantum effects enhance the breaking of the Lorentz symmetry towards large scales. On the other hand, the flow of the breaking couplings dies out quickly in the infrared since in this regime $g \propto k^2$ as $k\to0$, owing to the correct classical limit of a finite Newton's constant.

To decide which effect dominates, a numerical analysis has to be done. For this, we diagonalise the flow equations for the breaking couplings, \eqref{eq:lineqs}. The eigenvalues of the corresponding matrix are approximately $-0.21, -1.01 \pm 0.11\mathbf{i}$, thus the diagonalised couplings $d_1,d_2,d_3$ follow the flow
\begin{equation}
\begin{aligned}
 \dot d_1 &\approx -0.21 g \, d_1 \, , \\
 \dot d_{2,3} &\approx (-1.01 \pm 0.11\mathbf{i}) g \, d_{2,3} \, .
\end{aligned}
\end{equation}
These equations can be easily integrated. The infrared value of the coupling at scale $k$ is related to its value at some ultraviolet scale $\Lambda_\text{UV}$ by
\begin{equation}
\begin{aligned}
 d_{1,k} &\approx d_{1,\Lambda_\text{UV}} \exp\left[ 0.21 \int_k^{\Lambda_\text{UV}} \text{d}k \frac{g}{k}\right] \, , \\
 d_{2,3,k} &\approx d_{2,3,\Lambda_\text{UV}} \exp\left[ (1.01 \mp 0.11\mathbf{i}) \int_k^{\Lambda_\text{UV}} \text{d}k \frac{g}{k}\right] \, .
\end{aligned}
\end{equation}
Ignoring the oscillatory behaviour due to the complex part of the eigenvalues, the largest magnification of the Lorentz breaking is in the couplings $d_{2,3}$.

\begin{figure}[t]
\includegraphics[width=\columnwidth]{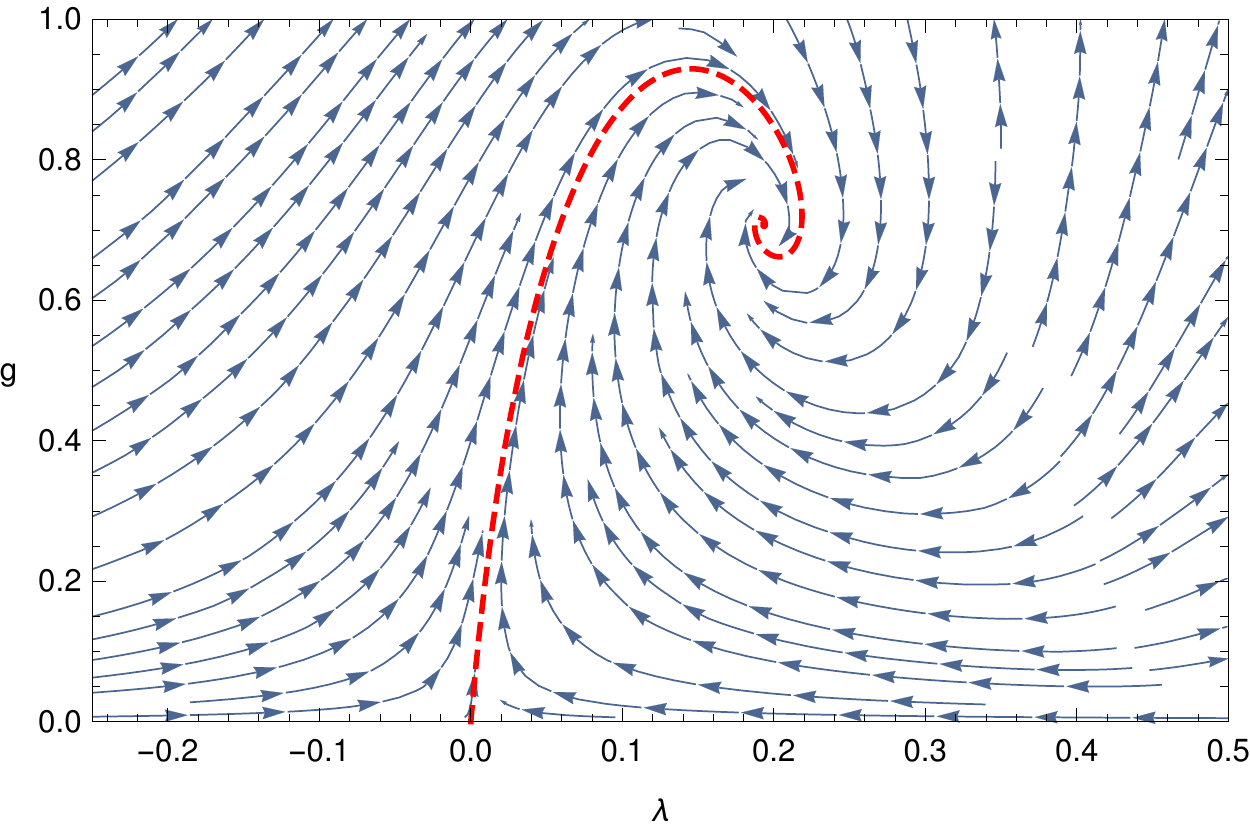}
\caption{Phase diagram of the diffeomorphism invariant cut plane. Arrows point towards higher energies. The red, dashed trajectory is the separatrix joining the ultraviolet fixed point and the Gaussian fixed point.}
\label{fig:phasediag}
\end{figure}

Let us quantify the magnification factor. First, it is clear due to our invariant setting that the original fixed point of the Einstein-Hilbert truncation persists, with breaking couplings set to zero. This is the only nontrivial fixed point in our approximation because of the linearisation in the breaking couplings. From this it is clear that we cannot sensibly take the limit $\Lambda_\text{UV}\to\infty$ to evaluate the magnification factor, since $d_{i,\Lambda_\text{UV}} \to 0$ in that limit, whereas the exponential diverges. Let us nevertheless try to give an estimate for the magnification.
For this, we consider the separatrix connecting the Gaussian and the nontrivial fixed point, as seen in \autoref{fig:phasediag}, which is close to the trajectory realised in nature. For scales below the Planck mass $M_\text{Pl}$, Newton's constant runs canonically to a very good approximation, $g \approx \frac{1}{M_\text{Pl}} k^2$. We will take the Planck scale to cut off the scale integral:
\begin{equation}
 \int_0^{\sqrt{M_\text{Pl}}} dk \frac{g}{k} = \frac{1}{M_\text{Pl}} \int_0^{\sqrt{M_\text{Pl}}} dk \, k = \frac{1}{2} \, .
\end{equation}
With this, we can estimate the magnification factors for the diagonalised couplings:
\begin{equation}
 d_{1,0} \approx 1.11 d_{1,{\sqrt{M_\text{Pl}}}} \, , \quad d_{2,3,0} \approx 1.66 d_{2,3,{\sqrt{M_\text{Pl}}}} \, .
\end{equation}
Translating back to the original couplings, we have
\begin{equation}
 \begin{pmatrix} k_{0,0} \\ k_{2,0} \\ a_{1,0} \end{pmatrix} \approx \begin{pmatrix} 1.15 & 0.01 & 0.22 \\ 0.01 & 1.35 & -0.24 \\ 0.24 & 0.52 & 1.92 \end{pmatrix} \begin{pmatrix} k_{0,{\sqrt{M_\text{Pl}}}} \\ k_{2,{\sqrt{M_\text{Pl}}}} \\ a_{1,{\sqrt{M_\text{Pl}}}} \end{pmatrix} \, .
\end{equation}
This indicates that during the flow the breaking couplings increase by a factor of order unity. Thus in practice even though Lorentz symmetry breaking is relevant, Lorentz symmetry breaking quantum gravity theories stand a chance if they provide a mechanism that drives the breaking couplings close to zero in the trans-Planckian regime. On the other hand, the calculation also shows that even if Lorentz symmetry breaking would be irrelevant, chances are that the breaking is not washed out in the infrared, because the flow of Newton's constant strongly suppresses the flow of the breaking couplings.

To conclude, the very stringent experimental bounds on Lorentz violations together with the present results make a Lorentz symmetry breaking theory of quantum gravity less attractive, since the constraints are essentially scale-independent up to very high scales, pointing towards a substantial amount of fine-tuning.

\section{Conclusions and Outlook}

In this letter, we have put forward several results:
\begin{itemize}
 \item We have introduced a renormalisation group equation for foliated spacetimes which preserves diffeomorphism symmetry if the original action is diffeomorphism invariant. For foliated spacetimes, there might exist a well-defined Wick rotation, our flow thus could provide a direct link of Euclidean flows to flows on foliatable Lorentzian manifolds.
 \item The new flow equation allows for a systematic study of renormalisation group flows of terms which break diffeomorphism symmetry. In particular, this should yield a very close link between the continuum approach and the Monte Carlo simulations of \CDT{}, where an anisotropy parameter is introduced. This connection shall be investigated in the future.
 \item As an application of the new flow, we derived the leading order flow equations of all couplings arising in a foliated setup, including up to two derivatives on the quantum fields. The analysis of this flow shows that generically the flow is not attracted towards the subspace spanned by diffeomorphism invariant action functionals. That means that General Relativity does \emph{not} emerge from the \RG{} running of \HL{} gravity. The numerical analysis shows that the enhancement of the breaking is rather small, thus Lorentz symmetry breaking theories are not ruled out entirely by our analysis if they have a mechanism to drive the breaking couplings close to zero already at the Planck scale. The very slow \RG{} running implies that the constraints on Lorentz symmetry violations essentially hold also at the Planck scale, posing significant challenges to any Lorentz symmetry breaking theory of quantum gravity.
\end{itemize}
A crucial point in a full description of nature is clearly the addition of matter, together with the corresponding Lorentz breaking terms. These might have the potential to change the relevance of the symmetry breaking. Furthermore the addition of matter allows to disentangle different speeds of light, which is one of the key signatures searched for in experiments on Lorentz symmetry breaking.

\section*{Acknowledgements}

I would like to thank Stefan Lippoldt, Chris Ripken and Frank Saueressig for interesting discussions, and Stefan Lippoldt and Frank Saueressig for critical comments on the manuscript.
I also profited from the fruitful discussion during the Mainz Institute for Theoretical Physics (MITP) workshop ``Quantum Fields - From Fundamental Concepts to Phenomenological Questions'', where the results of this letter were presented.
This research is supported by the Netherlands Organisation for Scientific Research (NWO) within
the Foundation for Fundamental Research on Matter (FOM) grant 13VP12.

\bibliographystyle{elsarticle-num}
\bibliography{general_bib}

\end{document}